# New insight into the dynamics of rhodopsin photoisomerization from one-dimensional quantum-classical modeling

Alexey S. Shigaev

*Institute of Mathematical Problems of Biology, Russian Academy of Sciences, Vitkevich st.1, 142290, Pushchino, Moscow region, Russia*

Tatiana B. Feldman[*]

*Biological Faculty, Lomonosov Moscow State University, Leninskie Gory1, 119991, Moscow, Russia;*
*Emanuel Institute of Biochemical Physics, Russian Academy of Sciences, Kosygin st.4, 119334, Moscow, Russia*

Victor A. Nadtochenko

*Semenov Institute of Chemical Physics, Russian Academy of Sciences, Kosygin st.4, 119991, Moscow Russia*

Mikhail A. Ostrovsky

*Biological Faculty, Lomonosov Moscow State University, Leninskie Gory1, 119991, Moscow, Russia;*
*Emanuel Institute of Biochemical Physics, Russian Academy of Sciences, Kosygin st.4, 119334, Moscow, Russia*

Victor D. Lakhno

*Institute of Mathematical Problems of Biology, Russian Academy of Sciences, Vitkevich st.1, 142290, Pushchino, Moscow region, Russia*

[*]Author to whom correspondence should be addressed; E-Mail: feldmantb@mail.ru; Tel.: +7-499-135-70-73; Fax: +7-499-137-41-01.

**ABSTRACT**

Characterization of the primary events involved in the *cis-trans* photoisomerization of the rhodopsin retinal chromophore was approximated by a minimum one-dimensional quantum-classical model. The developed mathematical model is identical to that obtained using conventional quantum-classical approaches, and multiparametric quantum-chemical or molecular dynamics (MD) computations were not required. The quantum subsystem of the model includes three electronic states for rhodopsin: (i) the ground state, (ii) the excited state, and (iii) the primary photoproduct in the ground state. The resultant model is in perfect agreement with experimental data in terms of the quantum yield, the time required to reach the conical intersection and to complete the quantum evolution, the range of the characteristic low frequencies active within the primary events of the 11-*cis* retinal isomerization, and the coherent character of the photoreaction. An effective redistribution of excess energy between the vibration modes of rhodopsin was revealed by analysis of the dissipation process. The results confirm the validity of the minimal model, despite its one-dimensional character. The fundamental nature of the photoreaction was therefore demonstrated using a minimum mathematical model for the first time.

**Keywords:** Visual pigment rhodopsin; Retinal chromophore; Dynamics of c*is-trans* photoisomerization; Quantum-classical model

# I. INTRODUCTION

The visual rhodopsin is a G-protein-coupled receptor involved in light perception information transfer [1-3]. Rhodopsin consists of a seven transmembrane alpha-helical apoprotein that can covalently bind the 11-*cis* retinal chromophore to $Lys_{296}$ in the seventh helix. Light quantum absorption leads to 11-*cis* retinal isomerization to the all-*trans* form, and this chromophore photoisomerization induces conformational changes in the protein that result in the formation of intermediates with various lifetimes and spectral properties. Finally, rhodopsin phototransformation results in hydrolysis of the Schiff base linkage and the release of all-*trans* retinal.

Chromophore photoisomerization is the first and only photochemical reaction in the complex process of phototransduction. The photoreaction is characterized by unique parameters. The primary event in chromophore photoisomerization takes place over 80–100 fs [4-6] and with a quantum yield of 0.67 [7], resulting in the formation of the primary ground-state rhodopsin photoproduct within 200 fs [8-11]. This reaction exhibits coherent character [11-16], with the coherent wave packet formed following femtosecond pulse impact. Wave packet dynamics can be observed using the absorption signals of both the excited and the ground states of the photoreaction products, and the coherence relaxation time is approximately 1 ps.

For the rhodopsin photochemical reaction, a 2-state model with a barrierless $S_1$ potential energy surface [11,17] and the involvement of a conical intersection [4-6,16,18-22] is generally accepted (Fig. 1). The time taken to reach the conical intersection is estimated to be about 80 fs [4-6]. Recently, however, resonant ultrafast heterodyne-detected transient-grating spectroscopy experiments led to the proposal that the photoproduct is formed predominantly during a single coherent event in a ~30 fs timescale [12]. In addition, the excited-state lifetime estimated from the low quantum yield of the weak spontaneous emission of rhodopsin is $50 \pm 20$ fs [23]. All these data indicate a probable excited-state lifetime in the range of 30–80 fs.

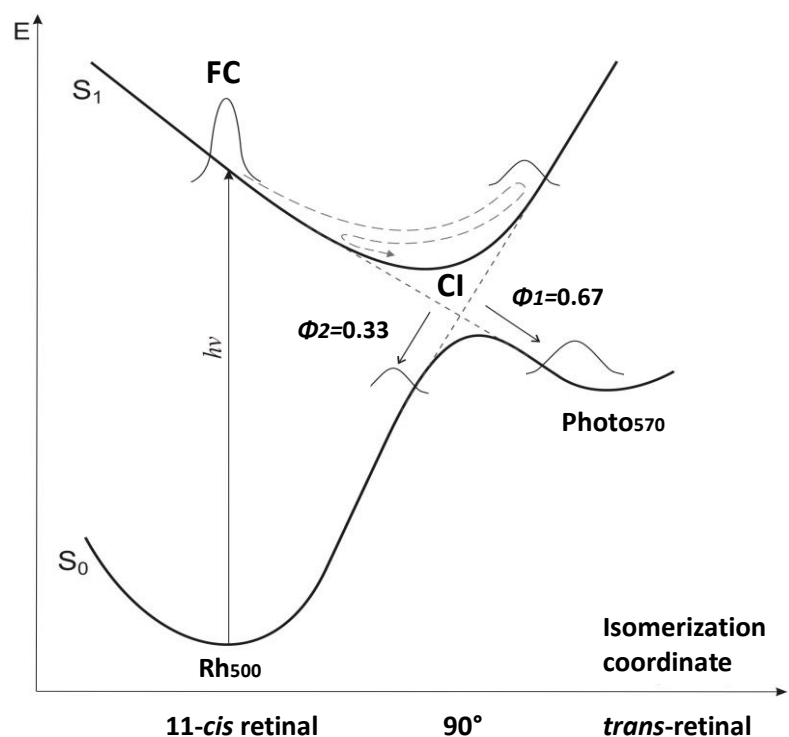

**FIG. 1.** Hypothetical scheme showing the potential energy surfaces participating in the photoreaction of rhodopsin. Evolution of the wave packet, which migrates along the $S_1$ potential energy surface from the initial Franck-Condon (FC) state to the $S_0/S_1$ region on the conical intersection (CI), is shown. The transition from the $S_1$ to the $S_0$ potential energy surface in the vicinity of the intersection point corresponds to reduction to the initial state ($Rh_{500}$) and formation of the primary ground-state photoproduct ($Photo_{570}$), with quantum yields of 0.33 and 0.67, respectively.

Considerable theoretical and experimental work has been performed in an attempt to explain the mechanism of the photoreaction [5,17-20,24-30]. The ultrafast rate observed for retinal chromophore photoisomerization indicates that perturbation caused by the photoreaction is initially localized at a relatively small region of the chromophore, specifically around the $C_{11}=C_{12}$ double bond [5,18,31-36]. First, the $\beta$-ionone ring of retinal is assumed to be strongly fixed through strong electrostatic interactions with amino acid residues $Trp_{265}$, $Phe_{261}$, and $Tyr_{268}$ of the opsin aromatic cluster [37-40]. Second, the tail of the retinal residue is bound covalently to $Lys_{296}$ and can be assumed to be motionless at the picosecond timescale. Consequently, the atomic groups that are movable as an entire entity during the photoreaction must be of a moderate molecular weight. Moreover, the coherent character of the photoreaction [11-16] also indicates the participation of a relatively small number of atoms.

One of the most effective methods of investigation of complex systems like rhodopsin is the multiscale modeling: calculation of the energy of an enzyme is performed by combining molecular mechanics modeling of the environment with quantum chemical modeling of the core region in which the chemically interesting action takes place [30,41]. Nevertheless, the abovementioned details of the rhodopsin chromofore center structure allow to propose a minimum mathematical model for the retinal chromophore photoisomerization in the femtosecond timescale. The successful approximation of photoisomerization reaction by a minimum model would be, in turn, a good supplementary evidence of elementary character of this photoreaction. Such a model should be insightful, should not constitute an excessively large computational burden, and should be able to account for key features and reproduce experimental data accurately.

In the present work, we developed a one-dimensional quantum-classical model of retinal chromophore photoisomerization at the femtosecond timescale. The resultant model is comparable with conventional quantum-classical models that have been successfully applied in theoretical studies of physical and chemical systems since 1959 [42-56]. We subsequently used our model to investigate the primary events involved in rhodopsin photoisomerization. Our investigation of the quantum-classical model for retinal *cis-trans* photoisomerization pursued two major objectives. The first goal was to reduce the characterization of the primary events of the *cis-trans* photoisomerization to a minimal one-dimensional quantum-classical model, because a valid one-dimensional approach would provide complementary evidence for the fundamental nature of conformational changes in the retinal chromophore during the photoreaction. As mentioned above, good agreement with experimental data is paramount, and key model parameters must be within physically reasonable ranges. The second objective was to clarify the nature of the dynamics of the primary photoisomerization process at the femtosecond timescale. Model parameters can be separated into two groups; the first can be estimated from the actual physical parameters of the modeled system, and the second consists of adjustable parameters whose ranges can only be estimated from the model and from comparison with experimental data. For example, interpretation of the friction coefficient range (which takes account of the mechanical impedance of the apoprotein) provides valuable insight into the nature of the excess energy dissipation and its relationship with quantum evolution (see section 3.2).

Our quantum-classical model of retinal photoisomerization was investigated using a wide range of parameters, and calculations revealed the best agreement with experimental data when key parameters of the model were close to the most physically realistic values. The calculations confirmed the primary events and localized nature of the conformational changes of the retinal chromophore. Fast redistribution of the excess vibrational energy suggests that the friction coefficient is important. Indeed, effective dissipation of excess energy was shown to be essential for agreement between quantum evolution dynamics and experimental data. Additionally, the model could successfully reproduce the experimentally observed coherent character of the photoreaction and the low-frequency fluctuations of the photoproduct backbone.

## II. THE MODEL

### A. Development of the quantum-classical model

Quantum-classical approaches play an important role in theoretical studies of various physicochemical systems [42-56]. Such approaches have recently proved to be powerful for investigating complex quantum-mechanical processes in biological systems, including charge separation in the photosynthetic reaction center of bacteria [49,50], and charge migration in homopolymeric [51-53] and heteropolymeric DNA [54-56].

The principal physics of quantum-classical approaches is the fact that the sum of nuclear masses of a molecule is greater than the sum of electron masses at least by four orders of magnitude. Therefore, the system of electrons may be described with the discrete Schrödinger equation, whereas the set of nuclei may be described with classical (Newtonian) mechanics equations. The experience of calculations shows (since 1959 [42]) that the approximation gives quite precise results.

It is of importance to note that both a single atom (nucleus) and an atomic group (set of nuclei) may be described as a mass point in the classical mechanics equations of a quantum-classical model. The mass points have certain equilibrium positions, corresponding to its energy minima. The displacement of the mass point from its equilibrium position is regulated by the elastic constant. The value of the elastic constant depends on properties of correspondent chemical bonds, see below. The thermal fluctuations may be introduced into the classic subsystem by the Langevin equation, realization of collisional thermostat, etc.

We propose that rhodopsin transitions through the series of conformational states during the photoreaction are analogous to the aforementioned charge migration in DNA and charge separation in the bacterial photosynthetic reaction center. Therefore, it may be described by an analogous minimum quantum-classical model. As shown below, this assumption allowed the adoption of an approach that proved to be in good agreement with all currently available experimental data on the rhodopsin retinal chromophore photoreaction.

To model the process of retinal chromophore photoisomerization in rhodopsin, a system involving three electronic states was chosen ("electronic" state is essentially vibronic ones, but in context of quantum-classical approaches they hereafter traditionally called as "electronic"):

1) $S_{0Rh}$ – the ground state with the retinal chromophore in the 11-*cis* conformation

2) $S_{1Rh}$ – the photoexcited state

3) $S_{0Photo}$ – the ground state of the primary photoproduct with the retinal chromophore in the *trans*-form.

These three electronic states comprise a quantum subsystem of the model with indexes 0, 1, and $X$ standing for $S_{0Rh}$, $S_{1Rh}$, and $S_{0Photo}$, correspondingly. Analogous indexes stand for all corresponding variables and parameters of model: 0 – for the ground state with the retinal chromophore in the 11-*cis* conformation, 1 – for the photoexcited state of the rhodopsin molecule, and $X$ – for the primary photoproduct with the retinal chromophore in the *trans*-conformation. The Hamiltonian of chromophore photoisomerization can be represented as follows:

$$H = \sum_n \nu_n |\nu_n\rangle\langle\nu_n| + \sum_{n \neq k} \nu_{nk} |\nu_n\rangle\langle\nu_k| \qquad (1)$$

where $n = 0, 1$, or $X$; $\nu_n$ is the energy of the $n^{\text{th}}$ electronic state [$eV$] with the wave function $|\nu_n\rangle$, and $\nu_{nk}$ are transfer matrix elements [$eV$] from the $n$-state to the $k$-state, $\langle\xi| = |\xi\rangle*$. To describe the excitation transfer, we can solve the Schrödinger equation:

$$i\hbar \frac{\partial}{\partial \tilde{t}} |\Psi\rangle = H |\Psi\rangle \qquad (2)$$

with the wave function in the form

$$|\Psi\rangle = \sum_n b_n(\tilde{t}) |n\rangle \qquad (3)$$

Here, the term $b_n(\tilde{t})|n\rangle$ is the amplitude of the probability of the occurrence of the system at the $n^{\text{th}}$ electronic state. At any moment, the normalization condition is fulfilled where $|b_0|^2 + |b_1|^2 + |b_X|^2 = 1$.

We assumed conventionally that the energy of the $n^{\text{th}}$ electronic state is a linear function of $\tilde{u}_k$ – the displacements [Å] of the correspondent mass points from their equilibrium positions, given by

$$\nu_n = \nu_n^0 + \sum_k \alpha'_{nk} \tilde{u}_k \qquad (4)$$

where $\nu_n^0$ is the energy of the $n^{\text{th}}$ electronic state under the condition $\tilde{u}_k = 0$ and $\alpha'_{nk}$ [$eV\cdot Å^{-1}$] are the electron-vibration coupling constants. The extent of the displacement is regulated by the elastic constants $K_k$ [$eV\cdot Å^{-2}$], see common description of quantum-classical approaches above. For simplicity, in this work, $K_0 = K_1 = K_X = K$. Coordinates $\tilde{u}_0$ and $\tilde{u}_X$, as well as elastic constants $K_0$ and $K_X$, correspond to the displacement of the ground state with the retinal chromophore in the 11-*cis* conformation, and the primary photoproduct with the retinal chromophore in the *trans*-conformation, respectively. Coordinate $\tilde{u}_1$ and constant $K_1$ corresponds to the displacement of the photoexcited state of the rhodopsin molecule.

Herein, we focused on using the masses $M_n$ ($M_0$, $M_1$, and $M_X$) as effective model parameters. These are not related to any particular atomic groups of the retinal chromophore, and $M_0 = M_1 = M_X = M$ [$kg$]. Obviously, the same rationale was applied for the effective elastic constants $K_n$ and the electron-vibration coupling constants $\alpha'_{nk}$. As well, all $\alpha'_{nk}$ are equivalent to the generalized electron-vibration coupling constant $\alpha'$. This approximation may appear crude at first glance, but it is in good agreement with the coherent character of the photoreaction [11-16]. Indeed, according to our previous experimental data, the coherent wave packet includes only one or two major low frequencies [14,15]. Consequently, it is conceivable that only one frequency may be specified here; hence $\tilde{\omega} = (K \cdot M^{-1})^{1/2}$ [$s^{-1}$] (see below).

Substituting expressions (1), (3), and (4) into the Schrödinger equation (2) and taking into account that $(v_{ij}) = (v_{ij})^*$, $i \neq j$ (i.e., all these coefficients are real numbers), we obtain the following:

$$i\hbar \dot{b}_n = v_n^0 b_n + \sum_k \alpha'_{nk} \tilde{u}_k b_n + \sum_{j \neq n} v_{jn} b_j \qquad n = 1, 2, 3 \qquad (5)$$

The complete Hamiltonian $H'$ of the system under consideration, averaged by the $|\Psi\rangle$ state, has the form:

$$\langle \Psi | H' | \Psi \rangle = \frac{M \dot{\tilde{u}}_n^2}{2} + \frac{K \tilde{u}_n^2}{2} + \langle \Psi | H | \Psi \rangle \qquad (6)$$

The motion equations for this Hamiltonian have the form:

$$M \ddot{\tilde{u}}_n = -K \tilde{u}_n - \gamma \dot{\tilde{u}}_n - \sum_n \alpha' |b_n|^2 \qquad (7)$$

where $M$ – the generalized mass of the mass points, $K$ – the generalized elastic constant, $\alpha'$ – the generalized electron-vibration coupling constant and $\gamma$ is an effective friction coefficient [$N \cdot s \cdot m^{-1} = kg \cdot s^{-1}$]. In the absence of thermal fluctuations friction coefficient $\gamma$ turns into a "phenomenological remainder" from the Langevin equation, but this parameter is virtually most important for the model investigation. The friction coefficient $\gamma$ takes account both of the mechanical impedance of the apoprotein and of the energy dissipation processes: collisional dissipation, transfer into apoprotein part and redistribution between vibration modes of the retinal chromophore during decoherence process, see below. First, we confined ourselves to the case of contact interaction where $\alpha'_{nk} = \delta_{nk}\alpha'_n$ ($\delta_{nk}$ – Kronecker sign). Second, we denoted the energy of the $n^{th}$ electronic state $v_n^0$ under the condition $\tilde{u}_k = 0$ (see expressions (4) and (5)) as $v_n$ definitely to remove all superscripts.

Thus, the developed quantum-classical model has the following form:

$$\begin{aligned}
\hbar i\, (db_0/d\tilde{t}) &= v_{01}b_1 + v_{0X}b_X + \alpha'\tilde{u}_0 b_0 - v_0 b_0 \\
\hbar i\, (db_1/d\tilde{t}) &= v_{1X}b_X + v_{01}b_0 + \alpha'\tilde{u}_1 b_1 - v_1 b_1 \\
\hbar i\, (db_X/d\tilde{t}) &= v_{1X}b_1 + v_{0X}b_0 + \alpha'\tilde{u}_X b_X - v_X b_X \\
M(d^2\tilde{u}_0/d\tilde{t}^2) &= -K\tilde{u}_0 - \gamma(d\tilde{u}_0/d\tilde{t}) - \alpha' |b_0|^2 \\
M(d^2\tilde{u}_1/d\tilde{t}^2) &= -K\tilde{u}_1 - \gamma(d\tilde{u}_1/d\tilde{t}) - \alpha' |b_1|^2 \\
M(d^2\tilde{u}_X/d\tilde{t}^2) &= -K\tilde{u}_X - \gamma(d\tilde{u}_X/d\tilde{t}) - \alpha' |b_X|^2
\end{aligned} \quad (8)$$

where $v_{01}$, $v_{0X}$, and $v_{1X}$ are non-diagonal matrix elements of the transition between the states in the quantum subsystem [$eV$]. Since $v_0$, $v_1$, and $v_X$ are diagonal matrix elements, corresponding to the energies of states $S_{0Rh}$, $S_{1Rh}$, and $S_{0Photo}$, they were assumed to be equal to the absolute values of energy state separation between $v_1$ and $v_n$; hence $v_X = |v_1 - v_X|$, $v_0 = |v_1 - v_0|$, and $v_1 = 0$.

From system (8), it follows that the probabilistic characteristics $|b_n|^2$ specify the real electrostatic strengths in the model (a common trait of the quantum-classical models). Therefore, large displacements in the atomic groups of the retinal chromophore could reveal a weak point in our approach. Fortunately, first, these displacements are small, as described previously [35]. Second, owing to the one-dimensional character of our model, the origins of the $\tilde{u}_0$, $\tilde{u}_1$, and $\tilde{u}_X$ coordinates may be related to any points of the rhodopsin chromophore center. Moreover, the corresponding coordinate axes may be situated at any angles relative to each other up to 180°. Third, it is important to realize that the real equilibrium positions of the 11-*cis* retinal and the primary photoproduct do not coincide with the points $\tilde{u}_0 = 0$ and $\tilde{u}_X = 0$. These positions correspond to the conditions $K\tilde{u}_n = -\alpha' |b_n|^2$ that occur immediately after completion of quantum subsystem evolution. Essentially, the displacements $\tilde{u}_n$ are the effective model parameters like $M$, $K$, or $\alpha'$.

### B. Assigning values for parameters

In this investigation, the non-diagonal matrix element $v_{0X}$ was set as 0 to forbid transfer between the two ground states $S_{0Rh}$ and $S_{0Photo}$. The values of matrix elements $v_0$ and $v_X$ can be easily calculated from the wavelength of light absorbed by the quantum states $S_{0Rh}$ and $S_{0Photo}$, and were 500 nm for $S_{0Rh}$ and 570 nm for $S_{0Photo}$. Consequently, $v_0 = 2.481$ eV and $v_X = 2.17$ eV. Nevertheless, it is difficult to identify even approximate ranges for the non-diagonal matrix elements $v_{01}$ and $v_{1X}$.

A similar situation is observed for the effective electron-vibration coupling constant $\alpha'$. The total displacement of atomic groups in the retinal chromophore during photoisomerization is very small (~1 Å) [35]. Taking into account the corresponding value of $v_0$ and (most likely) the nonlinear character of the dependence of $\alpha'$ on $\tilde{u}_n$, we specified the range $1 \leq \alpha' \leq 4$ eV·Å$^{-1}$.

The key values when choosing classical subsystem parameters are the effective mass $M$ and the characteristic time $\tau$ of the system. The latter was chosen to be equal to $10^{-15}$ s in view of the femtosecond timescale of retinal chromophore photoisomerization.

Although the effective mass is not related to any specific atomic group of the retinal chromophore, it may be estimated from the retinal structure. The quantum-classical model assumes that perturbation caused by the photoreaction in the picosecond timescale is initially localized on a relatively small part of the chromophore, specifically the $C_{11}=C_{12}$ double bond [5,18,31-36]. In our approximation, the $\beta$-ionone ring of retinal was assumed to be fixed (see section I), as was the chromophore tail. The $C_{11}=C_{12}$ double bond can be assumed to be the origin of coordinates, allowing arbitrary axes to be drawn over the $C_7$–$C_{12}$ and $C_{11}$–$C_{15}$ regions. The effective mass $M$ can then be set as the molecular weight of the most prominent part about these axes.

Through this approach, the effective mass was assumed to be equal to 27 a.m.u. (the sum of the molecular weights of the $C_{20}$ retinal methyl group and the $C_{13}$ atom). Additionally, this value is numerically equal to the sum of the molecular weights of the retinal $C_{19}$ methyl group and the $C_9$ atom. The corresponding atomic groups are shown in Fig. 2. The effective mass value was chosen based on the available data that indicate the active participation of the $C_{20}$ methyl group in the primary event of retinal photoisomerization [57-61].

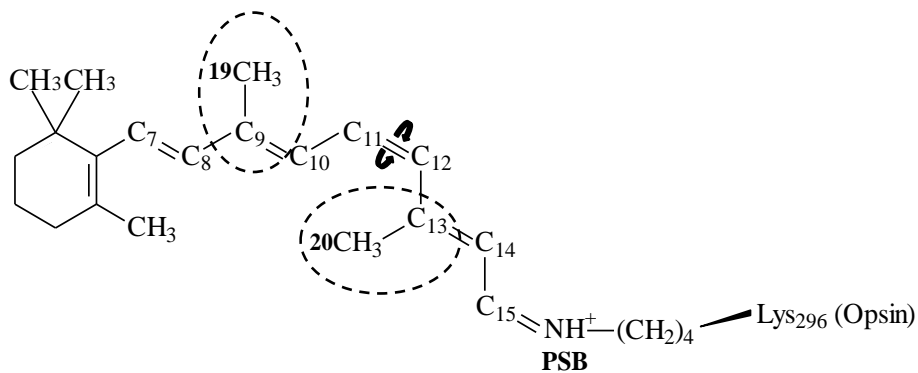

**FIG. 2.** Structure of the 11-*cis* retinal chromophore covalently bound to $Lys_{296}$ of the rhodopsin apoprotein via a protonated Schiff base (PSB) linkage. Atomic groups with a molecular weight of 27 a.m.u. (effective mass *M*, see text) are enclosed in ovals. Curved arrows around the $C_{11}=C_{12}$ bond identify the location of the 11-*cis* retinal chromophore during photoisomerization to the *trans*-form.

It is difficult to assess the effective range of the elastic constant *K* values explicitly. Nevertheless, the effective range of the retinal chromophore atomic groups vibration frequency $\tilde{\nu} = (K \cdot M^{-1})^{1/2} \cdot (2\pi)^{-1} [s^{-1}]$ may be readily estimated from numerous experimental data on the characteristic low-frequency vibration modes of the rhodopsin chromophore in different states [12,13,28,62,63]. These data cover a wide range of relevant low-frequency modes between 60 and 568 $cm^{-1}$.

Next, we considered the classical subsystem as a damped harmonic oscillator, and our test calculations confirmed that the resulting circular frequency $\tilde{\omega}_R$ of the classical subsystem oscillations was only slightly dependent on the quantum subsystem parameters. Hence, it can be safely assumed that $\tilde{\omega}_R$ depends only on $\tilde{\omega}$ and the friction coefficient $\gamma$ (see eq. 8) based on the well-known physical law $\tilde{\omega}_R = 2\pi\tilde{\nu}_R = (\tilde{\omega}^2 - 0.25\gamma^2)^{1/2}$. We primarily specified the range of $\tilde{\nu}_R$ from 110 to 220 $cm^{-1}$ using data from molecular dynamics (MD) studies [28], whereas $\tilde{\omega}^2$ values were calculated as the sums of corresponding $\tilde{\omega}_R^2$ and $0.25\gamma^2$ values.

The friction coefficient $\gamma$ is a key adjustable parameter of the model. This value takes account of the mechanical impedance of the apoprotein and the dissipation processes. Classical subsystem properties define the quantum subsystem evolution dynamics; hence a key feature of quantum-classical approaches is the ability to study the physical basis of a variety of physicochemical processes. To investigate this inter-relation for the primary photoreaction of the retinal chromophore using our model, we specified a very wide range for the friction coefficient $\gamma$, from $0.1 \times 10^{-12}$ to $6 \times 10^{-12}$ N·m$^{-1}$·s.

The lower value of the $\gamma$ range was estimated from the optimal friction coefficient in the quantum-classical Peyrard-Bishop-Holstein model for DNA ($0.6 \times 10^{-12}$ N·m$^{-1}$·s; see [55,64]). The friction coefficient in the Peyrard-Bishop-Holstein model takes into account the high rigidity of the DNA sugar-phosphate backbone, the presence of an ion coat made from phosphates, and the large weight of hydrated DNA bases. The lower limit of $\gamma$ in our approach therefore had to be significantly less than this. The upper limit of $\gamma$ was chosen to be 10-fold larger than the typical $\gamma$ value in the Peyrard-Bishop-Holstein model. The specified ranges of $\tilde{\omega}_R$ and $\gamma$ correspond to a range of $\tilde{\nu}$ from 110 to 418 cm$^{-1}$. Clearly, this interval is within the experimental $\tilde{\nu}$ range described above.

### C. Non-dimensionalization of the motion equations system

To transform system (8) into a dimensionless form, the following condition was specified:

$$\alpha' \tau^2 / M\, U = 1 \qquad (9)$$

where $U$ is an arbitrary scale of the displacement. It is evident that, in our approach, $U \approx \alpha' \cdot 0.0003574 \cdot 10^{-10}$ m·Å·eV$^{-1}$. The dimensionless time $t$ and the site displacement $u_n$ were determined by the following expression:

$$t = \tilde{t} / \tau; \quad u_n = \tilde{u}_n / U \qquad (10)$$

The dimensionless parameters $\eta_n$ of the quantum subsystem relate to their dimensional forms as follows:

$$\eta_n = \nu_n \tau / \hbar \qquad (11)$$

where $n$ corresponds to 0, 1, $X$, 01, 0$X$, or 1$X$.

The dimensionless friction coefficient $\Omega$ and the mass point vibration frequency $\omega^2$ relate to their dimensional forms as follows:

$$\Omega = \gamma\tau / M; \quad \omega^2 = \widetilde{\omega}^2 \cdot \tau^2 = K\tau^2 / M \qquad (12)$$

Hence, the dimensionless form of the resulting circular frequency is given by:

$$\omega_R^2 = \widetilde{\omega}_R^2 \tau^2 = (\widetilde{\omega}^2 - 0.25\gamma^2)\tau^2 = \omega^2 - 0.25\Omega^2 \qquad (13)$$

The electron-vibration coupling constant was transformed into the corresponding dimensionless parameter using the following expression (see also eq. 9):

$$\kappa = (\alpha')^2 \tau / \hbar K \quad \rightarrow \quad \kappa\omega^2 = \alpha' U\tau / \hbar \qquad (14)$$

The product $\kappa\omega^2$ is the dimensionless form of the electron-vibration coupling constant $\alpha'$ (the parameter $\kappa$ itself does not have a physical meaning). Thus, the dimensionless motion equations are as follows:

$$\begin{aligned}
i\dot{b}_0 &= \eta_{01}b_1 + \eta_{0X}b_X + u_0 b_0 \kappa\omega^2 - \eta_0 b_0 \\
i\dot{b}_1 &= \eta_{1X}b_X + \eta_{01}b_0 + u_1 b_1 \kappa\omega^2 - \eta_1 b_1 \\
i\dot{b}_X &= \eta_{1X}b_1 + \eta_{0X}b_0 + u_X b_X \kappa\omega^2 - \eta_X b_X \\
\ddot{u}_0 &= -\omega^2 u_0 - \Omega\dot{u}_0 - |b_0|^2 \\
\ddot{u}_1 &= -\omega^2 u_1 - \Omega\dot{u}_1 - |b_1|^2 \\
\ddot{u}_X &= -\omega^2 u_X - \Omega\dot{u}_X - |b_X|^2
\end{aligned} \qquad (15)$$

### D. Method of calculation and parameter sets

System (15) was solved numerically by the Runge-Kutta method to the fourth order of accuracy with a constant integration step. The accuracy of the solution was controlled by verifying at each step of the $||b_0|^2 + |b_1|^2 + |b_X|^2 - 1| < \varepsilon$ normalization condition. The value $\varepsilon$ was chosen to be equal to $10^{-4}$. Furthermore, calculations were made for different integration step values, and the resulting solutions were compared. Numerical investigation of system (15) was conducted with the initial conditions $|b_1|^2 = 1$, $|b_0|^2 = |b_X|^2 = 0$; $u_1 = u_X = 0$.

The first objective of the calculations was to obtain the loci of the model parameters when the model behavior is as close to the experimental data as possible. The second objective was to investigate the behavior of $\gamma$ ($\Omega$) as a function of all other parameters within these loci.

The electron-vibration coupling constant $\alpha'$ with values 1, 2, 3, and 4 eV·Å$^{-1}$ gave corresponding $\kappa\omega^2$ values of 0.000543, 0.002172, 0.004887, and 0.008688 dimensionless units. Values of the friction coefficient $\gamma$ multiplied by $10^{-12}$ were 0.1, 0.2, 0.3, 0.4, 0.5, 0.6, 1.2, 1.8, 2.4, 3.0, 3.6, 4.2, 4.8, 5.4, and 6.0 N·m$^{-1}$·s. The effective oscillation frequency $\widetilde{\omega}_R$ of the classical subsystem was varied from 110 to 220 cm$^{-1}$ with a difference interval of 10 cm$^{-1}$. The $\omega^2$ values were calculated for all combinations of the specified frequency $\widetilde{\omega}_R$ and the friction coefficient $\gamma^2$ as $\omega^2 = (\widetilde{\omega}_R^2 + 0.25\gamma^2)\tau^2 = \omega_R^2 + 0.25\Omega^2$ (see eq. 12 and 13). Thus, 720 combinations of $\kappa\omega^2$, $\omega^2$, and $\Omega$ were analyzed.

Values of non-diagonal matrix elements $\eta_{1X}$ and $\eta_{01}$ were varied from 0 to 0.256 dimensionless units with a difference interval of 0.002. Consequently, for each of the 720 combinations [$\kappa\omega^2$, $\omega^2$, $\Omega$] ([$\alpha'$, $\widetilde{\omega}^2$, $\gamma$]), 16,384 combinations of $\eta_{1X}$ and $\eta_{01}$ values were analyzed.

### III. RESULTS AND DISCUSSION

#### A. Evaluating model performance using experimental data

A calculation was performed for the first 500 fs after the moment of the retinal chromophore photoexcitation. To compare the model with experimental data, five performance indicators were introduced into the calculation.

The first indicator was the residual population of the $S_{1Rh}$ state, and further testing of the calculation for all parameters was performed only when the condition $|b_1|^2 < 0.005$ was fulfilled within the 475 to 500 fs time window.

The second indicator was the quantum yield $\Phi$ of the primary photoproduct formation ($S_{Photo}$), defined as a mean value of $|b_X|^2$ in the 475 to 500 fs time window. The experimentally measured quantum yield of retinal photoisomerization in rhodopsin is 0.67 [7], compared with 0.6–0.75 calculated using the model.

The third indicator was the time taken to reach conical intersection, defined as the time needed to reach the condition $|b_X|^2 > \Phi \pm 0.01$ for the first time. According to recent experimental data, this was between 30 [12] and 70–80 fs [4-6,23] (see above). We therefore attributed a value of 50 ± 20 fs.

The fourth indicator was the time taken to complete the evolution of the quantum subsystem. The final population of the $S_{1Rh}$ state was calculated as the mean $|b_1|^2$ value in the 475 to 500 fs time window for the result of each calculation. Then, the time-averaged population of the $S_{1Rh}$ state $|b_1(t)|^2$ for each time interval from $t$ to $(t + 25 \text{ fs})$ was divided by the final population of the $S_{1Rh}$ state $|b_1(475)|^2$. The terminal time for quantum subsystem evolution was defined as the moment the relation $(|b_1(t)|^2/(|b_1(475)|^2 = 1.02$ was reached for the first time. The experimentally measured time taken for the photoexcited molecule to transition from the $S_{1Rh}$ potential energy surface to the ground-state energy level of the primary photoproduct ($S_{0Photo}$) was 110–125 fs [4-6]. The time of the transition from the excited state to the primary photoproduct ground state was therefore specified as 100–130 fs.

The fifth indicator was the characteristic range of the resulting frequencies during the photoreaction. Every value of parameter $\Omega$ has its own set of frequency $\omega^2$ values specified so that the condition $110 \leq \tilde{\nu} \leq 220 \text{ cm}^{-1}$ is fulfilled (see above). In this case, combinations of $\eta_{1X}$ and $\eta_{01}$ in good agreement with experiment data were consistently in narrower ranges of $\omega^2$ that depend on $\kappa\omega^2$ and $\Omega$ (see below). The experimentally measured low-frequency fluctuations of the retinal chromophore throughout its photoisomerization were 136, 149, and 156 cm$^{-1}$ [11,12,14,15,65].

The photoisomerization of retinal in rhodopsin is a vibrationally coherent photochemical process. Thus, an additional comparison test would ideally be used to assess the coherent character of the photoreaction. However, it is very difficult to quantify the agreement between coherence from model calculations and experimental data; hence only qualitative characteristics were considered. This additional comparison test can be considered as the sixth performance indicator.

## B. Results from model calculations

Three peculiarities must be considered to understand the behavior of the model. First, the full transition from $S_{1Rh}$ to $S_{0Rh}$ and $S_{0Photo}$ states requires knowledge of certain relationships between $\eta_{1X}$, $\eta_{01}$, $\eta_X$, $\eta_0$, $\omega^2$, $\Omega$, and $\kappa\omega^2$. Specifically, the quantum transition starts when particular resonances of $u_n$, $du_n/dt$, and quantum subsystem parameters are reached. Otherwise, the excited state $S_{1Rh}$ will have an infinite lifetime in the model. Alternatively, the quantum subsystem will jump between $S_{1Rh}$, $S_{0Rh}$, and $S_{0Photo}$ states with high amplitude and moderate frequency during the infinite lifetime. Thus, the condition $0.6 \leq \Phi \leq 0.75$ will never be fulfilled (see the first and second indicators described above). Second, the final three equations in system (8) are analogous to the motion equation describing a springed mass point. This becomes apparent after replacing the final term in the equation by the force of gravity. Third, the test calculations showed that the coherent character of the photoreaction in our model takes place only if $\eta_{1X} > \eta_{01}$. The results for various parameter combinations were summarized in Table 1.

For $\alpha' = 1$ eV·Å$^{-1}$, no photoreaction was observed in the model system. When $\alpha' = 2$ eV·Å$^{-1}$ ($\kappa\omega^2 = 0.002172$), the resonance of quantum and classical subsystem parameters yielded relatively small values of $\omega^2$ and $\Omega$, and the range of the 'effective' dimensionless friction $\Omega$ was 0.002–0.013. The upper limit of the $\Omega$ range corresponds to $\gamma = 0.6 \times 10^{-12}$ N·m$^{-1}$·s, which is equal to the effective friction in the Peyrard-Bishop-Holstein model for DNA [55,64]. The damping of the classical subsystem fluctuations is very slow; when $\Omega < 0.013$, the quantum subsystem is unstable throughout the 500 fs time interval, and jumps between states and/or undergoes large fluctuations in $|b_n|^2$ values.

**Table 1.** The model behavior for the different parameter combinations.*

| Conditions | | | Comparisons | | | |
|---|---|---|---|---|---|---|
| $\alpha'$ | $\gamma \times 10^{-12}$ N·m$^{-1}$·s | $\|b_1\|^2$ (t = 500 fs) | $\Phi$ | $\tau_{CI}$, fs | $\tau_{QE}$, fs | $\tilde{\nu}_R$, cm$^{-1}$ |
| 1 eV·Å$^{-1}$ | – | > 0.005 | – | – | – | – |
| 2 eV·Å$^{-1}$ | 0.1 | < 0.002 | 0.63 | ≈ 77 | ≈ 140 | 110 – 130 |
| 3 eV·Å$^{-1}$ | 1.2 | < 0.0007 | 0.66 | 52.8 – 57.2 | 100 – 138 | 120 – 160 |
| 4 eV·Å$^{-1}$ | 1.2 | < 0.0005 | 0.67 | 36 – 38.5 | 105 – 130 | 130 – 160 |
| **Experiments** | | 0 | 0.67 [7] | 30 – 70 [4-6,12,23] | 110 – 125 [4-6] | 136 – 156 [11,12,14,15,65] |

* $\alpha'$ – the electron-vibration coupling constant; $\gamma$ – the friction coefficient providing the best agreement with experiments for given $\alpha'$; $|b_1|^2$ (t = 500 fs) – the residual population of the $S_{1Rh}$ state (the first performance indicator); $\Phi$ – quantum yield of the primary photoproduct formation state (the second performance indicator); $\tau_{CI}$ – the time taken to reach conical intersection (the third performance indicator); $\tau_{QE}$ – the time taken to complete the evolution of the quantum subsystem (the fourth performance indicator); $\tilde{\nu}_R$ – the characteristic range of the resulting frequencies, cm$^{-1}$ (the fifth performance indicator).

When $\alpha' = 2$ eV·Å$^{-1}$, the photoreaction is observed, but the quantum subsystem exhibits very unstable behavior. Fig. 3 presents the quantum transition for one of the model parameters ($\eta_{1X} = 0.1149$, $\eta_{01} = 0.0888$, $\omega^2 = 0.000601$, $\Omega = 0.002$, $\kappa\omega^2 = 0.002172$). The corresponding resulting frequency $\tilde{\nu}_R$ is equal to 130 cm$^{-1}$, and the friction coefficient $\gamma = 0.1 \times 10^{-12}$ N·m$^{-1}$·s. This is a very small value for $\gamma$, even for a moderate $\alpha'$. As a consequence, the classical subsystem undergoes large amplitude fluctuations over a long time period, as shown for $u_X(t)$ in Fig. 3(a). As shown in Fig. 3(b), the iterative resonances failed to rapidly stabilize, and appreciable jumps in $S_{0Photo}$ and $S_{0Photo}$ state populations were evident (the $S_{0Rh}$ population is not shown). Nevertheless, the necessary conditions for resonances were fulfilled; the residual population of the $S_{1Rh}$ state is less than 0.2%, indicating an almost complete quantum transition, as shown in Fig. 3(c).

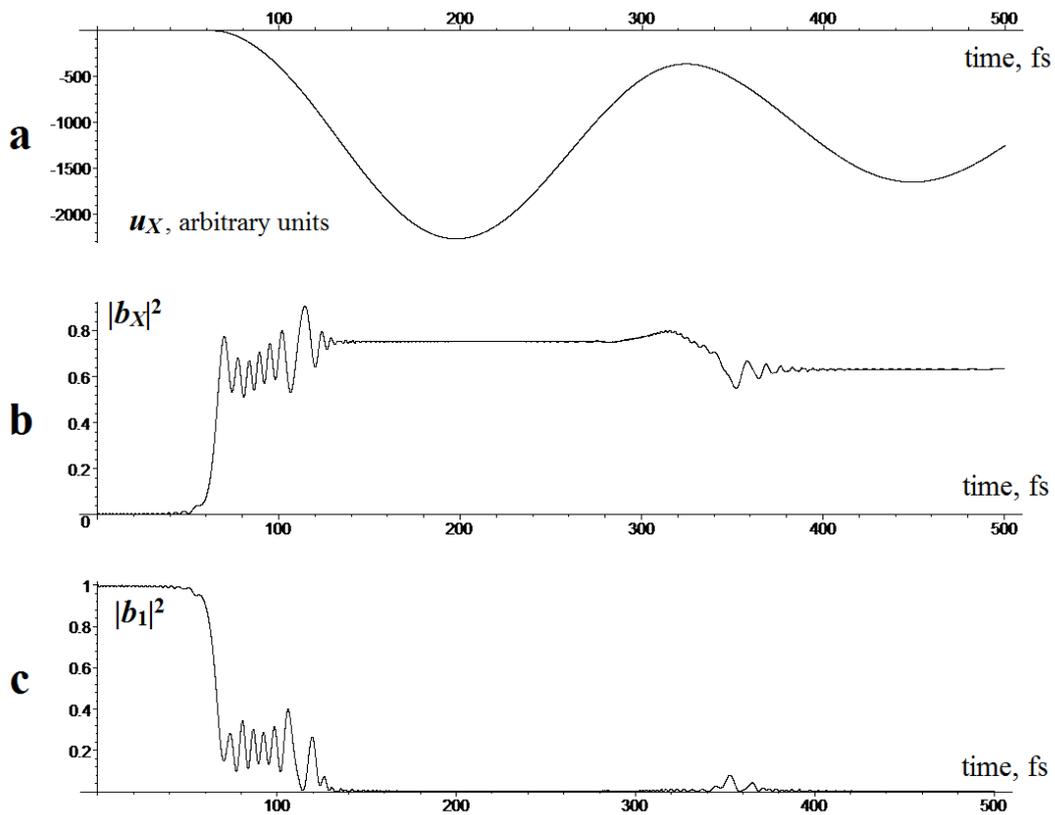

**FIG. 3.** Evolution of retinal quantum state populations following the moment of photoexcitation ($t = 0$ fs). **(a)** Evolution of the isomerization coordinate $u_X$ of the classical subsystem used to characterize the *cis-trans* transition of the photoexcited retinal chromophore. The maximum value of the characteristic displacement is 2268 a.u., corresponding to 1.62 Å. **(b)** Population of the $S_{0Photo}$ state. Large fluctuations of this value are clearly visible in the 270–400 fs time window. **(c)** Population of the $S_{1Rh}$ state. An almost complete quantum transition is clearly visible, but large jumps in $S_{1Rh}$ are also evident in the 270–400 fs time window. The final graph shows the transition of photoexcited rhodopsin molecules from the $S_1$ potential energy surface to the $S_0$ surfaces of $S_{0Rh}$ and $S_{0Photo}$ photoproducts.

When $\Omega = 0.013$, the system is sufficiently stable, but the time taken to reach the conical intersection is quite large (77 fs), and the characteristic time required to complete the evolution of the quantum subsystem is ~140 fs. The resulting $\tilde{\nu}$ frequencies are in the range of 110–130 cm$^{-1}$. Thus, when $\alpha' = 2$ eV·Å$^{-1}$, calculations from the model appear to be in good agreement with experimental data based on the first two indicators (the residual population of the $S_{1Rh}$ state and the quantum yield $\Phi$).

When $\alpha' = 3$ eV·Å$^{-1}$ ($\kappa\omega^2 = 0.004887$), an effective range for the friction constant is $0.004 \leq \Omega \leq 0.04$, with an upper limit corresponding to $\gamma = 1.8 \times 10^{-12}$ N·m$^{-1}$·s, which is three times larger than the characteristic value of the viscous friction in the Peyrard-Bishop-Holstein model [55,64]. The corresponding $\omega^2$ frequencies are also well above those when $\alpha' = 2$ eV·Å$^{-1}$. When $\varOmega \leq 0.13$, the model behaves stably over a narrow range of the resulting frequency (from 110 to 140 cm$^{-1}$). When $\varOmega \geq 0.027$, the upper limit of this range is 160 cm$^{-1}$. Thus, the range for the high friction coefficient includes all experimentally observed frequency values. When $\varOmega = 0.027$, the time taken to reach the conical intersection is 52.8–57.2 fs, and the time required to complete the evolution of the quantum subsystem is 100–138 fs. This is also the case when $\varOmega = 0.04$, but the resonance of quantum subsystem parameters with those of the classical subsystem and $\alpha'$ take place only when $\tilde{\nu} = 110$ cm$^{-1}$.

Hence, for $\alpha' = 3$ eV·Å$^{-1}$, model calculations are in good agreement with experimental data based on all five indicators, provided that $\varOmega = 0.027$.

For $\alpha' = 4$ eV·Å$^{-1}$ ($\kappa\omega^2 = 0.008688$), the range of effective $\varOmega$ values shifts to 0.007–0.054. Nevertheless, the quantum system exhibits quite unstable behavior up to $\varOmega \geq 0.013$. When $\varOmega$ is equal to 0.027 or 0.04, the quantum subsystem may be slightly unstable when $\tilde{\nu} > 180$ cm$^{-1}$. We analyzed the model behavior only for $\tilde{\nu}$ frequencies from 130 to 160 cm$^{-1}$. When $\varOmega = 0.027$, the time taken to reach the conical intersection is ~36–38.5 fs, and the time required to complete the evolution of the quantum subsystem is 105–130 fs. When $\varOmega = 0.04$, these ranges are 39–42.2 and 90–120 fs, respectively, and the typical time required for complete evolution of the quantum subsystem for $\varOmega = 0.054$ goes down to 80–90 fs.

Consequently, when $\alpha' = 4$ eV·Å$^{-1}$, model calculations are in good agreement with experimental data based on all five performance indicators only when $\varOmega = 0.027$, but the results obtained when $\varOmega = 0.04$ are also reasonable. These values of $\varOmega$ support the stable evolution of the quantum subsystem and rapid decay of classical subsystem oscillations.

The typical dynamics of the classical subsystem variable $u_X$ following the absorption of light by the retinal chromophore are presented in Fig. 4. The parameter values are $\eta_{1X} = 0.153$, $\eta_{01} = $ 0.1084, $\eta_{0X} = 0$, $\omega^2 = 0.000982$, $\Omega = 0.027$, and $\kappa\omega^2 = 0.008688$ ($\alpha' = 4$ eV·Å$^{-1}$), and the corresponding resulting frequency is 150 cm$^{-1}$.

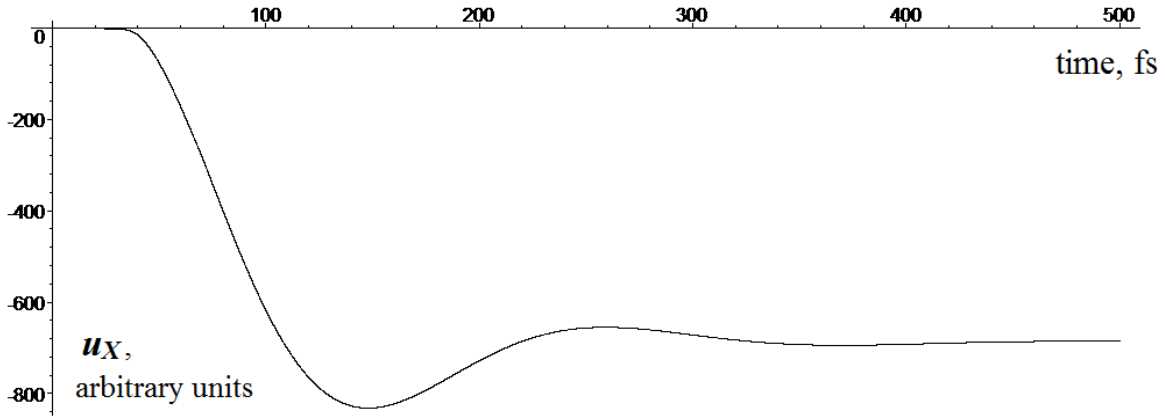

**FIG. 4.** Evolution of the isomerization coordinate $u_X$ for the classical subsystem with an optimal value for the viscous friction coefficient. The maximum value of the characteristic displacement is – 831 a.u., corresponding to 1.18 Å. The asymptotic value of $u_X$ at infinite time is approximately – 687 a.u. (measured for the 1000 fs interval), corresponding to ~0.98 Å.

As can be seen in Fig. 4, classical subsystem oscillations have a large rate of decay. A high value of the dimensionless viscous friction coefficient ($\Omega = 0.027$) corresponds to $\gamma = 1.2\times10^{-12}$ N·m$^{-1}$·s. This value is twice as high as the analogous parameter in the Peyrard-Bishop-Holstein model for DNA (see [55,64]). The corresponding dynamics of $S_{0Rh}$, $S_{1Rh}$, and $S_{0Photo}$ state populations within the first 500 fs after the absorption of a photon by the retinal chromophore are shown in Fig. 5.

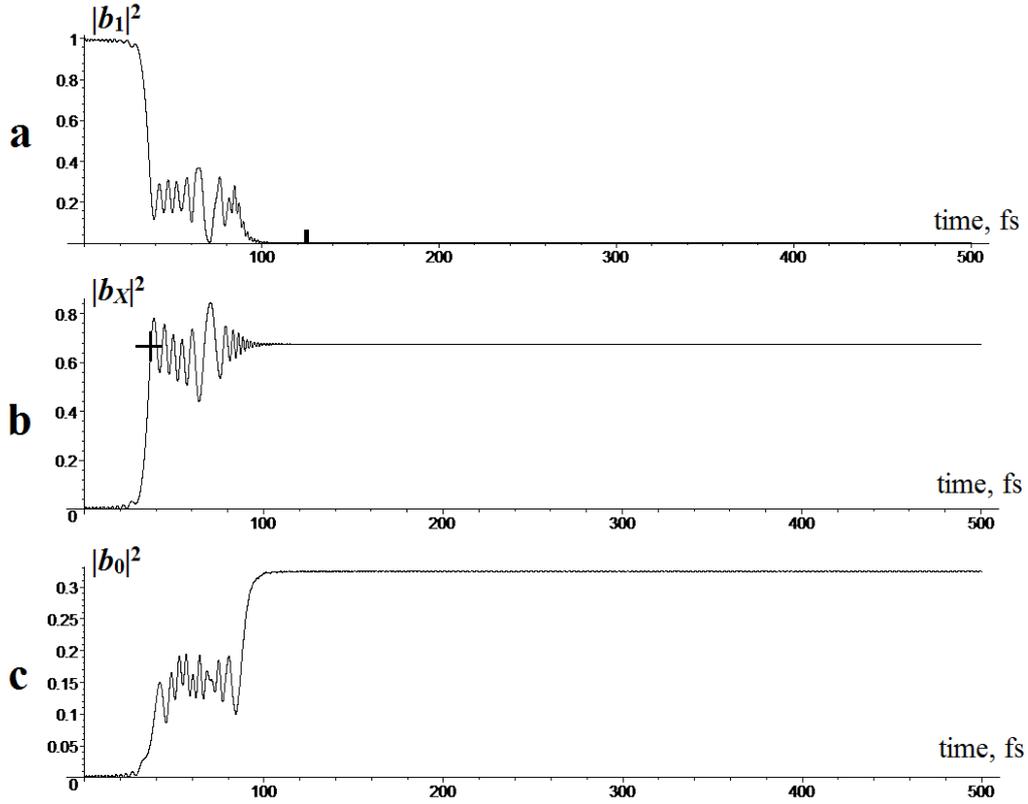

**FIG. 5.** Evolution of the quantum states of populations of the retinal chromophore following photoexcitation based on physically realistic model parameters. **(a)** The population of the $S_{1Rh}$ state. The time of the transition from the excited state to the primary photoproduct ground state (126 fs) is represented by the short vertical bar on the time axis. **(b)** Formation of the primary ground-state rhodopsin photoproduct $S_{0Photo}$ containing isomerized *trans*-retinal. The time taken to reach the conical intersection (CI) ($t = 37.2$ fs) is indicated by a cross. **(c)** Return of the photoexcited rhodopsin molecule to its initial state $S_{0Rh}$ containing non-isomerized 11-*cis* retinal. The population of the $S_{0Rh}$ state increases in two distinct steps (see main text for explanation).

Figures 4 and 5 show that rapid stabilization of the classical subsystem results in stable quantum evolution without jumps in populations of the electronic states (in contrast to Fig. 3). Nevertheless, the higher values of $\Omega$ give rise to classical subsystem overdamping and forbid quantum evolution.

The model does not allow elucidation of the relationship between collisional dissipation in the chromophore center and the redistribution of excess energy transfer between vibrational modes of rhodopsin. However, experimental data reveal a high vibrational excitation (T > 2000 K) for low-frequency modes of rhodopsin during photoisomerization [62], corresponding to the 282, 350, and 477 cm$^{-1}$ modes. First, these frequencies are much larger than the upper limit of the coherent wave packet frequency range (136–156 cm$^{-1}$), as discussed previously [11,12,14,15,65,66]. Second, the temperature maximum shown in Fig. 4 corresponds to 3280 K for 27 a.m.u., consistent with the report by Kim et al. [63]. It follows that the bulk of the excess energy is redistributed between vibrations of smaller atomic groups after the primary photoreaction (during the decoherence process). Moreover, this efficient redistribution may well be a key requirement for stable evolution of populations of electronic states and a high quantum yield.

The second topical issue is the coherent character of the photoreaction. This qualitative characteristic is the additional performance indicator listed above, and is very important for comparison of model-derived calculations and experimental data. It is difficult to assess the coherence effects quantitatively, but this feature of the model can clearly be grasped visually from the graph in Fig. 5 that shows the dynamics of the **S$_{0Rh}$**, **S$_{1Rh}$**, and **S$_{0Photo}$** states within the first 500 fs after absorption of a light quantum by the retinal chromophore. The values of dimensionless parameters are the same as those in the graph in Fig. 4.

It can be clearly seen that evolution of the quantum subsystem occurs in two phases. In the first phase, a sharp increase in the population of the primary photoproduct **S$_{0Photo}$** (b) and a slight increase in the population of the initial rhodopsin state **S$_{0Rh}$** (c) are observed. In the second phase (after the plateau in the 39–87 fs time window), the mean population of **S$_{0Photo}$** does not change, and the excited state **S$_{1Rh}$** passes only to **S$_{0Rh}$**. Such behavior is in good agreement with the conclusions on the coherent character of the retinal chromophore photoisomerization in rhodopsin based on experimental data [4,5,11-14,16]. The analogous behavior of quantum subsystem can be considered as coherence.

We analyzed this qualitative parameter for all combinations of $\kappa\omega^2$, $\omega^2$, and $\Omega$. The photoreaction has a very poorly resolved coherent character when $\alpha' = 2$ eV·Å$^{-1}$, but when $\alpha' = 3$ eV·Å$^{-1}$, the coherence of the quantum subsystem transition was very well-defined provided that $\Omega = 0.027$. When $\alpha' = 4$ eV·Å$^{-1}$, the coherent character of the photoreaction was pronounced only when $Ω = 0.027$. When $Ω = 0.04$, this was much less evident, and when $Ω = 0.054$ the coherence was barely observable.

Thus, $α′$ between 3 and 4 eV·Å$^{-1}$ and $Ω = 0.027$ proved to be optimal parameters and gave model-derived calculations in perfect agreement with experiment data based on all performance indicators, including the coherent character of the photoreaction. Moreover, departure of these key parameters from the above values results in the simultaneous loss in agreement based on all performance indicators apart from the first two.

Therefore, the model system behavior is in good agreement with experimental data only when model parameters are close to the most physically realistic values. First, this confirms the validity of our approach, and second, good agreement with experimental data using a minimal one-dimensional quantum-classical model illuminates the primary events and highly localized conformational changes during photoisomerization of the retinal chromophore.

The developed approach is easily expandable. For example, dependence of $α′$ on coordinates $u_n$ could be introduced to more precisely reproduce the behavior of the retinal chromophore during the photoreaction. Additionally, our model allows visualization of the characteristic low-frequency fluctuations of the retinal backbone that have been observed to occur after photoisomerization in various experiments [11,12,14,15,65,66]. Thus, our model provides novel insight into the photoisomerization process.

## IV. CONCLUSION

In summary, we showed that essential elements of the primary photoreaction of the rhodopsin retinal chromophore *cis-trans* photoisomerization are reducible and can be approximated by a minimum one-dimensional quantum-classical model. The developed mathematical model is identical to conventional quantum-classical approaches in that it does not require a large computational burden, and it provides important insight into the physical nature of various phenomena. Quantum-classical models have been successfully applied in theoretical studies of physicochemical systems for over 60 years.

The quantum subsystem of the model includes three electronic states of the rhodopsin molecule: the ground state $S_{0Rh}$, the excited state $S_{1Rh}$, and the primary photoproduct in the ground state $S_{0Photo}$. The classical subsystem includes three mass points with an equal mass of 27 a.m.u. that simulate regions of the retinal chromophore. The approach reproduces the dynamics of the non-adiabatic photochemical *cis-trans* photoisomerization of the retinal chromophore into the primary photoproduct.

A wide range of model parameters were investigated extensively, and model-derived calculations provide good agreement with experimental data only when key model parameters are close to the most physically realistic values. Six performance indicators were selected to judge model quality: the residual population of the $S_{1Rh}$ state, the quantum yield of the photoreaction, the time taken to reach the conical intersection, the time required to complete the quantum evolution, the characteristic range of low frequencies during the photoreaction, and the coherent character of the process.

These performance indicators confirmed the validity of the approach, despite its simplicity and one-dimensional character. Furthermore, the ability to accurately reproduce experimental data with such a minimal mathematical approach confirms the localized nature of the primary conformational changes at the retinal chromophore during photoisomerization. Thus, a valid minimum mathematical model provides theoretical evidence of the coherent nature, elementary character, and localization of the retinal molecular backbone displacements.

Analysis of the friction coefficient ranges suggests that the efficient redistribution of excess energy between different vibrational modes of rhodopsin in the photoreaction may well be a key requirement for stable evolution of populations of electronic states and achieving a high quantum yield. Consequently, the developed model holds promise for studying the mechanism of ultrafast photoisomerization of the retinal chromophore in other members of the large retinal-containing protein family.

## ACKNOWLEDGEMENTS

This work was supported by the Russian Foundation for Basic Research [No. 16-07-00305]; and the Russian Science Foundation [No. 16-11-10163]. The authors are thankful to the Joint Supercomputer Center of the Russian Academy of Sciences, Moscow, Russia for the provided computational resources.

## CONFLICT OF INTEREST

The authors declare that no conflict of interests regarding the publication of this paper arises.